# Deriving Analytical Solutions Using Symbolic Matrix Structural Analysis: Part 2 – Plane Trusses

Vagelis Plevris[1*] and Afaq Ahmad[2]

[1] College of Engineering, Qatar University
P.O. Box: 2713, Doha, Qatar
e-mail: vplevris@qu.edu.qa

[2] Department of Built Environment, Oslo Metropolitan University
P.O. Box 4, St. Olavs Plass, NO-0130, Oslo, Norway
e-mail: afahm2637@oslomet.no

*Corresponding author

**Abstract.** This study extends the use of symbolic computation in Matrix Structural Analysis (MSA) to plane (2D) trusses, building on previous work that focused on continuous beams. An open-source MATLAB program, hosted on GitHub, was developed to perform symbolic analysis of 2D trusses under point loads for any configuration. Using MATLAB's Symbolic Math Toolbox, the program derives analytical expressions for displacements, support reactions, and axial forces, providing a deeper understanding of truss behavior. The symbolic approach also supports efficient and scalable sensitivity analysis by directly computing partial derivatives of outputs with respect to input parameters, enhancing design exploration and optimization. This tool serves as a valuable resource for both engineering practice and education, offering clear insights into parameter relationships and enriching the understanding of structural mechanics. The accuracy of the symbolic results has been rigorously validated against two commercial finite element software programs and results from the literature, with full agreement, confirming the validity and generality of the methodology.

## 1 Introduction

The Finite Element Method (FEM), marking 80 years since its inception [1], is a cornerstone of structural analysis, providing a reliable numerical framework for solving complex engineering problems [2]. Traditional FEM, while effective for detailed results, is limited in flexibility and generality as numerical solutions are tied to specific inputs, requiring full re-computation for changes in material properties, geometry, or external loads [3]. Changes to these inputs require a full re-computation of the system, which is both time-consuming and computationally demanding, particularly for large-scale structures. Moreover, numerical results often conceal the relationships between key parameters, hindering a deeper understanding of structural behavior [4].

For linear structures like beams, trusses, and frames, FEM is often referred to as Matrix Structural Analysis (MSA), sharing its principles and origins. MSA for such linear structures allows direct derivation of stiffness matrices without numerical integration, simplifying the process and offering greater insight. A limitation of traditional numerical FEM and MSA is their dependence on predefined boundary conditions and loading scenarios. Any changes in the model or the loads require regenerating and recalculating the entire model, resulting in a repetitive and time-intensive workflow. This lack of adaptability restricts the use of these methods in tasks like real-time analysis or mathematical optimization [5], where evaluating multiple configurations quickly and efficiently is essential [6].

Before the advent of computers, most mathematical and engineering analyses relied heavily on closed formulas and symbolic computation, as numerical methods and solutions were virtually nonexistent. With the rise of computing technology, numerical methods rapidly evolved and became dominant in all scientific fields, offering efficient solutions to complex problems. However, efforts to bridge the gap between symbolic and numerical methods emerged, such as the development of the Macsyma system in the 1960s by MIT's AI group [7]. Symbolic computation can provide compelling solutions to challenges associated with numerical methods such as FEM and MSA. Unlike purely numerical methods, which produce specific values, symbolic computation enables the manipulation and solving of mathematical expressions in their exact algebraic form. This capability facilitates the derivation of analytical solutions and offers deeper insights into structural behavior, enhancing understanding and flexibility in the analysis.

MATLAB is widely used in structural engineering for its powerful numerical algorithms, addressing challenges related to matrix analysis of structures and FEM [8-11], structural dynamics [12-15], optimization [16-20], and others [21]. While renowned for numerical computing, MATLAB also supports symbolic computation through its Symbolic Math Toolbox [22, 23]. The toolbox extends the program's capabilities, enabling symbolic algebraic simplifications, differentiation, integration, equation solving, and matrix manipulation [24-26]. In engineering, physics, and mathematics, the toolbox is invaluable for analyzing complex systems, offering exact, parameterized solutions that enhance understanding and flexibility. For researchers and educators, it facilitates the derivation of closed-form solutions, exploration of theoretical concepts, and intuitive presentation of results. Its integration with MATLAB's numerical environment allows efficient transitions between symbolic and numerical analyses, providing a versatile platform for both theoretical studies and practical applications. Other similar numerical platforms also support symbolic computation, like Mathematica [27], Maple [28], and SymPy (Python) [29].

Building on previous work focused on continuous beams [4], this study extends the methodology to plane trusses, retaining shared features while introducing capabilities specifically designed for 2D truss analysis. The study presents an innovative, open-source MATLAB program tailored for symbolic MSA of 2D trusses subjected to joint point loads. For the first time, this program enables precise and efficient derivation of analytical solutions for any 2D truss configuration, offering symbolic expressions for node displacements, support reactions, and element axial forces. Key features of the program include:

- Fully open-source code, accessible online with comprehensive documentation and five illustrative numerical examples.

- Capability to derive closed-form solutions for various output quantities (e.g., displacements, support reactions, axial forces) for 2D trusses of any complexity.

- Support for sensitivity analysis, using MATLAB's built-in symbolic differentiation to assess the impact of input parameters on output quantities.

- Full validation of the symbolic results through comparison with established commercial finite element software and results from the literature, ensuring accuracy, reliability, and general applicability.

The source code, available on GitHub (https://github.com/vplevris/SymbolicMSA-2DTrusses), includes all examples discussed in this study. Developed using MATLAB R2024b and its symbolic toolbox, the program is expected to be compatible with earlier versions. With clean, well-documented MATLAB code and minimal setup required, users are encouraged to explore the program, customize it to suit their specific needs, and efficiently generate their own analytical solutions.

## 2 Literature Review

Few studies have addressed stiffness matrices symbolically. Eriksson and Pacoste [30] explored the use of symbolic software for developing finite element procedures, focusing on complex problems like higher-order instabilities requiring precise formulations. They emphasized that symbolic tools improve efficiency and clarity, enabling effective comparisons between element assumptions. Their research includes beam formulations for plane and space models, allowing analytical verification of equivalence between displacement and co-rotational approaches. Symbolic derivation also simplifies finite space rotations and systematically connects local displacements to global variables. Amberg et al. [31] developed a Maple-based toolbox for generating finite element codes, facilitating 1D, 2D, and 3D simulations. This toolbox has significantly accelerated research in areas like thermocapillary convection, welding, and crystal growth by reducing development time to hours. It offers flexibility, transparency, and ease of modification, enabling researchers to focus on physical insights while minimizing errors and debugging. Pavlovic [32] highlighted symbolic computation as a powerful complement to traditional numerical methods in structural engineering. He reviewed its underutilized applications and emphasized its potential to advance classical structural analysis by addressing complex problems with greater efficiency. He advocated for integrating symbolic and numerical methods to leverage their complementary strengths in solving structural mechanics problems effectively.

Symbolic algebra has found diverse applications in structural engineering, enabling precise and insightful formulations of complex problems. For example, Levy et al. [33] utilized symbolic algebra to derive geometric stiffness matrices for membrane shells, allowing for the consideration of finite rotations without relying on small rotation assumptions. This approach enhances nonlinear analysis by providing explicit and physically intuitive derivations, demonstrating the power of symbolic methods in advancing structural mechanics. Murphey [34] derived generic symbolic equations for the effective stiffness and strength of beam-like trusses with arbitrary numbers of longerons and diagonal lacings, assuming relatively soft diagonals. These equations unify previous discrete cases and simplify truss design by allowing modifications through constant values. Covering bending, torsion, shear, and axial loading, the approach enables rapid preliminary sizing and optimization, validated against finite element analysis and prior results by Renton [35].

Skrinar and Plibersek [36] derived a symbolic stiffness matrix and load vector for slender beams with transverse cracks under uniform loading. Using the principle of virtual work, they provided closed-form expressions that clarify the influence of crack depth and location on flexural deformation, aiding in crack identification and modeling per European design code EC8. Roque [37] explored symbolic and numerical analysis of bending plates using MATLAB, demonstrating its versatility in combining symbolic and numerical methods seamlessly to enhance problem-solving efficiency and accuracy.

Tinkov [38] derived new exact analytical expressions for the deflection of various planar trusses and analyzed existing solutions using the Maxwell-Mohr formula under elastic assumptions. Employing induction on the number of panels and symbolic computations in Maple, the study identified key characteristics and limitations related to panel count. Comparative analysis with known solutions, validated using the Lira software package, demonstrated the accuracy and applicability of the derived analytical expressions for truss deflection. Kirsanov and Tinkov [39] developed an algorithm to derive deflection and horizontal displacement formulas for planar statically determinate trusses under various loading conditions. Using symbolic mathematics in Maple, they generalized solutions for trusses with increasing panels, constructing and solving recurrence equations for polynomial coefficients. Their method identifies asymptotic properties and extreme points in deflection behavior, providing valuable benchmarks for validating numerical calculations in structural analysis. In a similar study, Kirsanov [40] proposed a rod model for planar statically determinate frames with four supports, deriving analytical deflection formulas under various loads using Maple. Through symbolic equilibrium equations, he identified kinematic instability for specific panel numbers and determined deflections for stable configurations using Maxwell-Mohr's formula. The method generalized solutions via double induction and recurrence equations, revealing significant deflection variations. Force expressions in critical rods were also derived, aiding design evaluation and optimization.

Dasgupta [41] explored the application of symbolic computations using Mathematica to derive equilibrium equations in terms of nodal displacements. Assuming linear elastic behavior and small displacements, the study demonstrates how bar stiffness matrices and nodal forces are integrated to construct the global system matrix. The Mathematica Solve function is employed to obtain exact solutions, offering an educational perspective on structural mechanics while showcasing the power of symbolic tools in finite element analysis. Öchsner and Makvandi [42] discuss the theory of single rod elements and plane truss structures, providing a detailed solution procedure through a comprehensive example. They combined manual calculations with insights into computer implementation using Maxima [43], enabling a clear understanding of the FEM. The work includes practical Maxima examples, facilitating the application of the methods to other problems and serving as a valuable resource for learning symbolic and numerical analysis in truss design.

In the area of structural optimization, analytical solutions continue to stand out, as demonstrated by Charalampakis and Chatzigiannelis [44]. They utilized the cylindrical algebraic decomposition (CAD) algorithm to derive globally optimal solutions for minimum weight truss design. Their methodology, supported by symbolic computation, provides exact solutions to benchmark problems, offering formal validation for the convergence of metaheuristic methods to global optima. While computationally intensive and not yet practical for real-life engineering, CAD's potential grows with advances in algorithmic improvements and computing power.

## 3   Stiffness Matrix of a Plane Truss Element

The stiffness matrix is essential in MSA as it establishes the relationship between applied forces and resulting displacements in a structural system. Typically, in MSA and the FEM, stiffness matrices are computed numerically for each element and then assembled into a global system representing the entire structure. However, for linear elements in MSA, it is often feasible to derive an exact symbolic expression for an element's stiffness matrix [42, 45].

One such example is the 2D Truss Element, which has one degree of freedom (DOF) per node in the element's local system (displacement along the element's axis). This element is commonly used in linear static analysis of plane trusses. Figure 1 shows the element in its local coordinate system, where the two DOFs—axial displacements at each end—are illustrated. In plane trusses, each node of the model has two DOFs in the global coordinate system, corresponding to displacements in the $x$ and $y$ global directions. There are no rotations; each element experiences only axial deformation, resulting in axial forces and stresses, with no shear forces or bending moments.

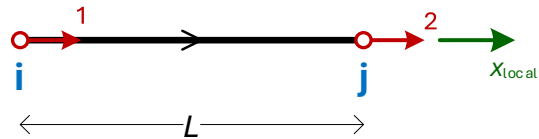

Figure 1. 2D Euler–Bernoulli beam element with 6 DOFs.

Since plane truss elements can have various orientations on the plane, each element's stiffness matrix must be transformed (rotated) to a common coordinate system to construct the global stiffness matrix of the whole model. Figure 2 shows an inclined element on the 2D plane, where the angle $\theta$ defines the rotation needed from the global $x$-axis to align with the local element axis with a counter-clockwise rotation. The figure also shows the local and global axes, with DOF numbering in the global system. For a 2D truss element, there are four DOFs in total, with two DOFs at each node, all referenced to the global axes, as shown in Figure 2.

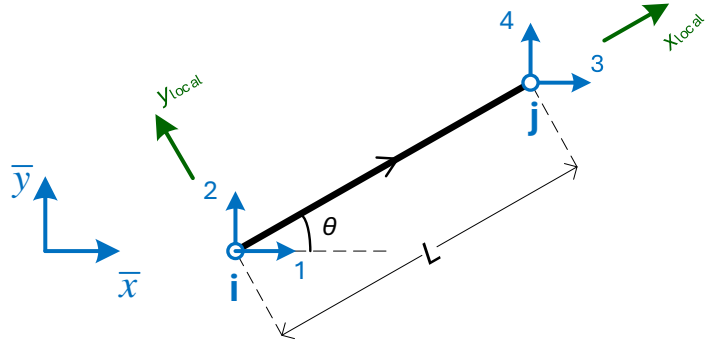

Figure 2. 2D truss element with 4 DOFs in the global system.

The symbolic 2×2 stiffness matrix for the 2D truss element, corresponding to the two local DOFs depicted in Figure 1, is expressed in Eq. (1).

$$[\hat{k}_{t2}] = \frac{EA}{L} \cdot \begin{bmatrix} 1 & -1 \\ -1 & 1 \end{bmatrix} \qquad (1)$$

The transformation matrix, which is a 2×4 matrix, is given by Eq. (2) [42]

$$[T_{t2}] = \begin{bmatrix} c & s & 0 & 0 \\ 0 & 0 & c & s \end{bmatrix} \qquad (2)$$

where:

$$c = \cos(\theta) = \frac{\Delta x}{L} = \frac{x_j - x_i}{L}$$
$$s = \sin(\theta) = \frac{\Delta y}{L} = \frac{y_j - y_i}{L} \qquad (3)$$

$$[T_{t2}] = \begin{bmatrix} c & s & 0 & 0 \\ 0 & 0 & c & s \end{bmatrix} \qquad (4)$$

The global stiffness matrix $[\bar{k}_{t2}]$ of the element, a 4×4 matrix corresponding to the four DOFs illustrated in Figure 2, is obtained using the following expression [42]:

$$[\bar{k}_{t2}] = [T_{t2}]^T \cdot [\hat{k}_{t2}] \cdot [T_{t2}] = \frac{EA}{L} \cdot \begin{bmatrix} c^2 & cs & -c^2 & -cs \\ cs & s^2 & -cs & -s^2 \\ -c^2 & -cs & c^2 & cs \\ -cs & -s^2 & cs & s^2 \end{bmatrix} \qquad (5)$$

In these equations, $E$ is Young's modulus of the material, $A$ is the cross-sectional area, and $L$ is the length of the truss element.

## 4 Definition of the Model Symbolically

In Part 1 of this work, which was focused on continuous beams [4], the beam model was defined symbolically using the variables: *Lengths*, *Supports*, *PointLoads*, and *UniformLoads*. This approach worked well for beams because all elements were aligned along a single line, making the structure relatively simple to describe using only the length of each element and assuming continuous numbering from left to right for both nodes and elements. However, in 2D trusses, the geometry of the model is more complex, with nodes that can connect multiple elements in various orientations. Therefore, in this 2D truss implementation, we introduce a different set of variables for our model, namely:

- *NodeCoords*: Coordinates ($x$, $y$) of the nodes (matrix of size *NumNodes*×2).

- *ElemMatSec*: The stiffness of each element, as the product $E \cdot A$, with $E$ being the modulus of elasticity of the material and $A$ the cross-sectional area (column vector of size *NumElements*).

- *ElemCon*: Connectivity of the elements, i.e. ID of start node $i$ and end node $j$ for each element (matrix of size *NumElements*×2).

- *Supports*: The supports for each node and each DOF, where 1 means constrained and 0 means free to move (matrix of size *NumNodes*×2).

- *PointLoads*: The point loads for each node and each DOF (matrix of size *NumNodes*×2).

As shown above, instead of specifying element lengths, we define each node's coordinates ($x$ and $y$ values) and specify the element connectivity, indicating which nodes each element connects. This setup allows the program to model any 2D truss geometry effectively, regardless of complexity, as the element lengths and orientations are automatically derived from the nodal coordinates. This approach provides greater flexibility, enabling efficient modeling of intricate truss structures without manually calculating individual element properties.

For instance, the input file for the first example, examined in Section 6, is as follows:

```
NodeCoords = [  0  0 ;
                L  L ;
              3*L  0];
ElemMatSec = [EA; EA];
ElemCon = [1 2 ;
           2 3];
Supports = [1 1 ;
            0 0 ;
            1 1];
PointLoads = [0  0 ;
              0 -P ;
              0  0];
```

## 5    Importance and Educational Benefits of Symbolic Solutions in 2D Truss Analysis

Symbolic representation in structural analysis offers engineers and researchers valuable insights into structural behavior. By preserving algebraic relationships between parameters, symbolic MSA enables a detailed exploration of how variations in material properties, geometry, or loading conditions influence the overall structural response. This approach provides parameterized solutions that are not limited to specific input values, offering flexibility and adaptability. Additionally, symbolic expressions allow for the straightforward calculation of partial derivatives with respect to input parameters, making sensitivity analysis and design optimization more efficient and accessible.

Symbolic MSA also holds significant educational value. It enhances the understanding of structural engineering concepts by clearly demonstrating the relationships between key parameters. Unlike purely numerical methods, which deliver results without exposing the underlying mechanics, symbolic solutions offer a transparent view of these interactions, fostering a deeper conceptual understanding. A key benefit of using symbolic solutions is the clear visualization of how parameters such as Young's modulus $E$, cross-sectional area $A$, and element length $L$ influence the nodal displacements or the axial force or stress in truss members. In contrast to numerical methods, which often give isolated numerical values, symbolic expressions explicitly show the dependencies between input parameters and the resulting displacements or internal forces. As a result, symbolic MSA is a powerful tool for research and also an effective resource for teaching and communicating complex structural mechanics in a clear and intuitive manner.

Symbolic solutions also help students grasp the concept of superposition and the cumulative effect of different loads on the truss system. By analyzing symbolic expressions for axial forces or nodal displacements, students can observe how various load cases contribute to the overall response, which may not be as evident in purely numerical results.

Another advantage has to do with sensitivity analysis and design optimization. Symbolic expressions allow for direct differentiation with respect to parameters such as $E$ or $A$, enabling students to easily explore how changes in material or geometric properties can affect the performance of the truss. This capability is crucial for understanding structural sensitivity and optimizing designs effectively.

In summary, symbolic analysis provides a transparent and accessible way to understand the relationships between structural parameters and the response of truss systems. It enhances learning by making complex concepts clearer and equips students with the skills needed for parametric studies, sensitivity analysis, and optimization. This approach offers a valuable educational tool for mastering the fundamentals of truss analysis in structural engineering.

## 6    Numerical Examples

We consider five numerical examples of differing levels of complexity. In these, $EA$ is treated as a single symbolic parameter, since $E$ and $A$ consistently appear together in stiffness and other analytical expressions in 2D trusses. Nevertheless, each member can have its own $EA$ properties, i.e. a member can be stiffer than another. Table 1 provides detailed descriptions of the numerical examples and the associated symbolic parameters for each.

Table 1. Details of the five numerical examples.

| Example # | Figure | Symbolic Parameters |
|-----------|--------|---------------------|
| 1 |  | 3 (*EA*, *L*, and *P*) |
| 2 |  | 4 (*EA*, *L*, *H*, and *P*) |
| 3 |  | 4 (*EA*, *L*, *H*, and *P*) |
| 4 |  | 4 (*EA*, *L*, *H*, and *P*) |
| 5 |  | 4 (*EA*, *L*, *H*, and *P*) |

## 6.1 Numerical Example 1 (3 Symbolic Parameters)

The first numerical example is a simple 2-bar truss with two members and a point load $P$, as shown in Figure 3. The symbolic parameters are three: $EA$, $L$, and $P$.

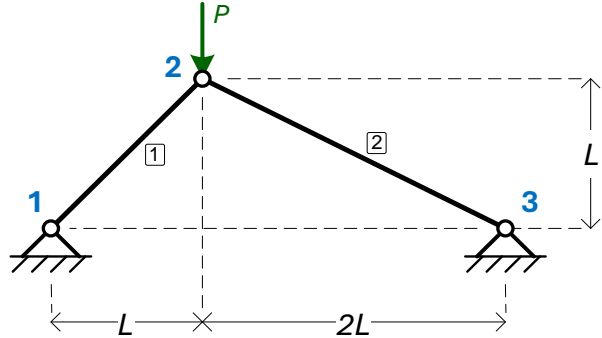

Figure 3. The truss of Example 1.

Table 2 shows the details of the model, as given in MATLAB. In every example, the number of nodes (*NumNodes*) is defined by the rows of the *NodeCoords* matrix and the number of elements (*NumElements*) is defined by the rows of the *ElemMatSec* matrix. In this example we have 2 elements and 3 nodes. In a 2D truss, all point loads are defined on nodes. There are no uniform loads.

Table 2. Details of the input parameters of the 1st numerical example.

| | |
|---|---|
| **NodeCoords** | $\begin{bmatrix} 0 & L & 3L \\ 0 & L & 0 \end{bmatrix}^T$ |
| **ElemMatSec** | $\begin{bmatrix} EA & EA \end{bmatrix}^T$ |
| **ElemCon** | $\begin{bmatrix} 1 & 2 \\ 2 & 3 \end{bmatrix}^T$ |
| **Supports** | $\begin{bmatrix} 1 & 0 & 1 \\ 1 & 0 & 1 \end{bmatrix}^T$ |
| **PointLoads** | $\begin{bmatrix} 0 & 0 & 0 \\ 0 & -P & 0 \end{bmatrix}^T$ |

Table 3, Table 4 and Table 5 present the results of the symbolic analysis in terms of the symbolic parameters. Table 3 presents the Node displacements, while Table 4 shows the support reactions and Table 5 the element axial forces. The element forces are reported here as positive when the member is in tension and negative when the member is under compression. The analytical expressions for the element stresses are not reported in the results, as the stress for any element can be easily found by dividing the force of the element with its cross-sectional area, $A$.

Table 3. Example 1: Node displacements.

| Node # | $x$-Displacement ($D_x$) | $y$-Displacement ($D_y$) |
|--------|--------------------------|--------------------------|
| Node 1 | 0 | 0 |
| Node 2 | $-\dfrac{PL\left(4\sqrt{2}-5\sqrt{5}\right)}{9EA}$ | $-\dfrac{PL\left(8\sqrt{2}+5\sqrt{5}\right)}{9EA}$ |
| Node 3 | 0 | 0 |

Table 4. Example 1: Support reactions.

| Node # | Force $F_x$ | Force $F_y$ |
|--------|-------------|-------------|
| Node 1 | $2P/3$ | $2P/3$ |
| Node 3 | $-2P/3$ | $P/3$ |

Table 5. Example 1: Element axial forces.

| Element # | Axial Force |
|-----------|-------------|
| 1 | $-\dfrac{2\sqrt{2}}{3}P$ |
| 2 | $-\dfrac{\sqrt{5}}{3}P$ |

## 6.2 Numerical Example 2 (4 Symbolic Parameters)

The second numerical example is the truss shown in Figure 4. It has 3 nodes and 3 elements. The symbolic parameters are four: $EA$, $L$, $H$, and $P$.

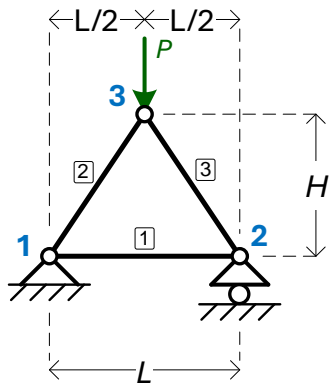

Figure 4. The truss of Example 2.

Table 6 shows the details of the model, as given in MATLAB.

Table 6. Details of the input parameters of the 2nd numerical example.

| | |
|---|---|
| **NodeCoords** | $\begin{bmatrix} 0 & L & L/2 \\ 0 & 0 & H \end{bmatrix}^T$ |
| **ElemMatSec** | $\begin{bmatrix} EA & EA & EA \end{bmatrix}^T$ |
| **ElemCon** | $\begin{bmatrix} 1 & 1 & 2 \\ 2 & 3 & 3 \end{bmatrix}^T$ |
| **Supports** | $\begin{bmatrix} 1 & 0 & 0 \\ 1 & 1 & 0 \end{bmatrix}^T$ |
| **PointLoads** | $\begin{bmatrix} 0 & 0 & 0 \\ 0 & 0 & -P \end{bmatrix}^T$ |

The results of the symbolic analysis are given in Table 7, Table 8, and Table 9, for Node displacements, support reactions, and element axial forces, respectively.

Table 7. Example 2: Node displacements.

| Node # | $x$-Displacement ($D_x$) | $y$-Displacement ($D_y$) |
|---|---|---|
| Node 1 | 0 | 0 |
| Node 2 | $\dfrac{PL^2}{4EAH}$ | 0 |
| Node 3 | $\dfrac{PL^2}{8EAH}$ | $-\dfrac{P\left((4H^2+L^2)^{3/2}+L^3\right)}{16EAH^2}$ |

Table 8. Example 2: Support reactions.

| Node # | Force $F_x$ | Force $F_y$ |
|---|---|---|
| Node 1 | 0 | $\dfrac{P}{2}$ |
| Node 3 | - | $\dfrac{P}{2}$ |

Table 9. Example 2: Element axial forces.

| Element # | Axial Force |
|---|---|
| 1 | $\dfrac{PL}{4H}$ |
| 2 | $-\dfrac{P\sqrt{4H^2+L^2}}{4H}$ |
| 3 | $-\dfrac{P\sqrt{4H^2+L^2}}{4H}$ |

### 6.3 Numerical Example 3 (4 Symbolic Parameters)

The third numerical example is the truss shown in Figure 4. It has 5 nodes and 7 elements. The symbolic parameters are four: $EA$, $L$, $H$, and $P$.

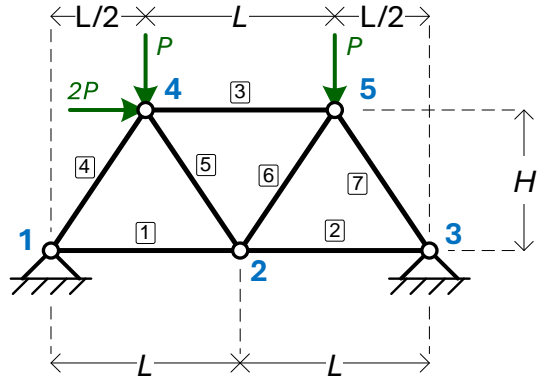

Figure 5. The truss of Example 3.

Table 6 shows the details of the model, as given in MATLAB.

Table 10. Details of the input parameters of the 3rd numerical example.

| | |
|---|---|
| **NodeCoords** | $\begin{bmatrix} 0 & L & 2L & L/2 & 3L/2 \\ 0 & 0 & 0 & H & H \end{bmatrix}^T$ |
| **ElemMatSec** | $\begin{bmatrix} EA & EA & EA & EA & EA & EA & EA \end{bmatrix}^T$ |
| **ElemCon** | $\begin{bmatrix} 1 & 2 & 4 & 1 & 2 & 2 & 3 \\ 2 & 3 & 5 & 4 & 4 & 5 & 5 \end{bmatrix}^T$ |
| **Supports** | $\begin{bmatrix} 1 & 0 & 1 & 0 & 0 \\ 1 & 0 & 1 & 0 & 0 \end{bmatrix}^T$ |
| **PointLoads** | $\begin{bmatrix} 0 & 0 & 0 & 2P & 0 \\ 0 & 0 & 0 & -P & -P \end{bmatrix}^T$ |

The results of the symbolic analysis are given in Table 11, Table 12, and Table 13, for Node displacements, support reactions, and element axial forces, respectively.

Table 11. Example 3: Node displacements.

| Node # | $x$-Displacement ($D_x$) | $y$-Displacement ($D_y$) |
|---|---|---|
| Node 1 | 0 | 0 |
| Node 2 | $\dfrac{PL}{2EA}$ | $-\dfrac{P\left(2HL^2 + \dfrac{(4H^2+L^2)^{3/2}}{2} + L^3\right)}{4EAH^2}$ |
| Node 3 | 0 | 0 |
| Node 4 | $\dfrac{P\left(H(4H^2+L^2)^{3/2} + 3HL^3 + L^4\right)}{4EAHL^2}$ | $-\dfrac{P\left(3HL^2 + (4H^2+L^2)^{3/2} + L^3\right)}{8EAH^2}$ |
| Node 5 | $-\dfrac{P\left(HL^3 - H(4H^2+L^2)^{3/2} + L^4\right)}{4EAHL^2}$ | $-\dfrac{P\left(HL^2 + (4H^2+L^2)^{3/2} + L^3\right)}{8*EA*H^2}$ |

Table 12. Example 3: Support reactions.

| Node # | Force $F_x$ | Force $F_y$ |
|---|---|---|
| Node 1 | $-\dfrac{\mathrm{P}(2H-L)}{2\mathrm{H}}$ | $-\dfrac{\mathrm{P}(H-L)}{L}$ |
| Node 3 | $-\dfrac{P(2H+L)}{2H}$ | $\dfrac{P(H+L)}{L}$ |

Table 13. Example 3: Element axial forces.

| Element # | Axial Force |
|---|---|
| 1 | $\dfrac{P}{2}$ |
| 2 | $-\dfrac{P}{2}$ |
| 3 | $-\dfrac{\mathrm{P}(2H+L)}{2H}$ |
| 4 | $\dfrac{P\sqrt{4H^2+L^2}\cdot(H-L)}{2HL}$ |
| 5 | $-\dfrac{P\sqrt{4H^2+L^2}}{2L}$ |
| 6 | $\dfrac{P\sqrt{4H^2+L^2}}{2L}$ |
| 7 | $-\dfrac{P\sqrt{4H^2+L^2}\cdot(H+L)}{2HL}$ |

## 6.4 Numerical Example 4 (4 Symbolic Parameters)

The fourth numerical example is the truss structure shown in Figure 6. The symbolic parameters are four: $EA$, $L$, $H$, and $P$.

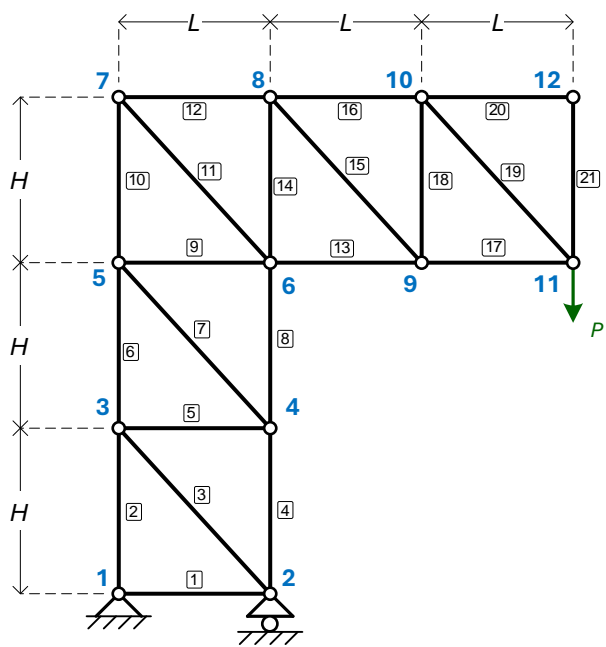

Figure 6. The truss of Example 4.

This model is relatively large, comprising 12 nodes and 21 elements. Detailed specifications of the model can be found in the input file for Example 4 in the source code repository. Each element is defined sequentially from the lower-numbered node to the higher-numbered node; for instance, element 7 connects nodes 4 and 5 in that order, while element 15 connects nodes 8 and 9. The supports include a pinned support at node 1 and a roller support at node 2, as illustrated in the figure.

The results of the symbolic analysis are reported in Table 14, Table 15 and Table 16, for Node displacements (for selected nodes); support reactions; and element axial forces (for all elements), respectively.

Table 14. Example 4: Node displacements (for selected nodes).

| Node # | $x$-Displacement ($D_x$) | $y$-Displacement ($D_y$) |
|:---:|:---:|:---:|
| Node 6 | $\dfrac{9H^2P}{AEL}$ | $-\dfrac{6HP}{AE}$ |
| Node 7 | $\dfrac{P\left(2(H^2+L^2)^{3/2}+21H^3\right)}{EAHL}$ | $\dfrac{6HP}{AE}$ |
| Node 9 | $\dfrac{P\left(9H^3-2L^3\right)}{EAHL}$ | $-\dfrac{P\left(3(H^2+L^2)^{3/2}+19H^3+4L^3\right)}{EAH^2}$ |
| Node 11 | $\dfrac{3P\left(3H^3-L^3\right)}{EAHL}$ | $-\dfrac{2P\left(3(H^2+L^2)^{3/2}+16H^3+5L^3\right)}{EAH^2}$ |

Table 15. Example 4: Support reactions.

| Node # | Force $F_x$ | Force $F_y$ |
|--------|-------------|-------------|
| Node 1 | 0 | $-2P$ |
| Node 2 | - | $3P$ |

Table 16. Example 4: Element axial forces.

| Element # | Axial Force | Element # | Axial Force |
|-----------|-------------|-----------|-------------|
| 1 | 0 | 11 | $-\dfrac{2P\sqrt{H^2+L^2}}{H}$ |
| 2 | $2P$ | 12 | $\dfrac{2LP}{H}$ |
| 3 | 0 | 13 | $-\dfrac{2LP}{H}$ |
| 4 | $-3P$ | 14 | $-P$ |
| 5 | 0 | 15 | $\dfrac{P\sqrt{H^2+L^2}}{H}$ |
| 6 | $2P$ | 16 | $\dfrac{LP}{H}$ |
| 7 | 0 | 17 | $-\dfrac{LP}{H}$ |
| 8 | $-3P$ | 18 | $-P$ |
| 9 | 0 | 19 | $\dfrac{P\sqrt{H^2+L^2}}{H}$ |
| 10 | $2P$ | 20 | 0 |
|   |   | 21 | 0 |

## 6.5 Numerical Example 5 (4 Symbolic Parameters)

The fifth numerical example is taken from the literature. In particular, it is the case No 1 truss model from the work of Tinkov [38]. The truss is shown in Figure 6. The symbolic parameters are four: $EA$, $L$, $H$, and $P$.

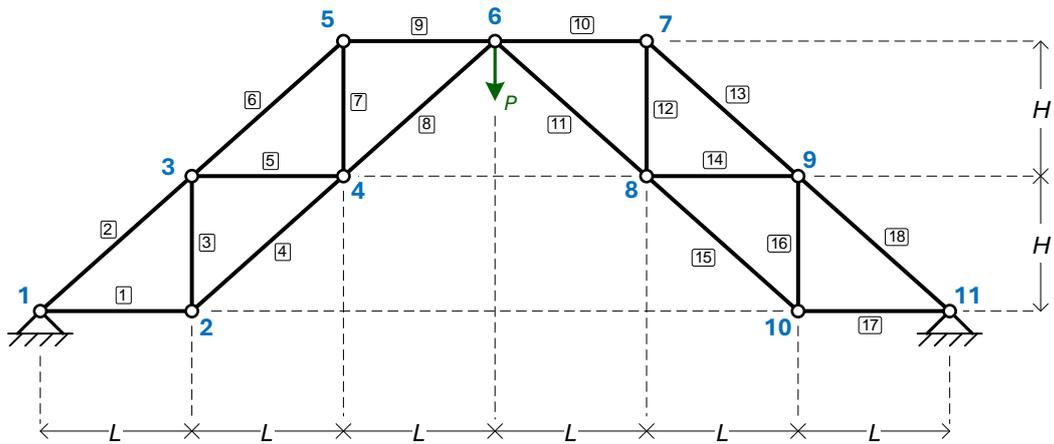

Figure 7. The truss of Example 5.

The model consists of 11 nodes and 18 elements, with detailed specifications available in the input file for Example 5 in the source code repository. Each element is defined sequentially, connecting the lower-numbered node to the higher-numbered node. The supports are pinned at nodes 1 and 11, as shown in the figure. The model is fully symmetric, including its loading, so identical results are expected on both sides of the truss.

The results of the symbolic analysis are given in Table 17, Table 18, and Table 19, for Node displacements (for selected nodes); support reactions; and element axial forces, respectively.

Table 17. Example 5: Node displacements (for selected nodes).

| Node # | $x$-Displacement ($D_x$) | $y$-Displacement ($D_y$) |
|---|---|---|
| Node 2 | $-\dfrac{L^2 P}{4EAH}$ | $-\dfrac{P\left(4(H^2+L^2)^{3/2}+2H^3+L^3\right)}{8EAH^2}$ |
| Node 4 | $-\dfrac{L^2 P}{8EAH}$ | $-\dfrac{P\left(3(H^2+L^2)^{3/2}+H^3+L^3\right)}{4EAH^2}$ |
| Node 6 | $0$ | $-\dfrac{P\left(10(H^2+L^2)^{3/2}+2H^3+3L^3\right)}{8EAH^2}$ |
| Node 8 | $\dfrac{L^2 P}{8EAH}$ | $-\dfrac{P\left(3(H^2+L^2)^{3/2}+H^3+L^3\right)}{4EAH^2}$ |
| Node 10 | $\dfrac{L^2 P}{4EAH}$ | $-\dfrac{P\left(4(H^2+L^2)^{3/2}+2H^3+L^3\right)}{8EAH^2}$ |

Table 18. Example 5: Support reactions.

| Node # | Force $F_x$ | Force $F_y$ |
|---|---|---|
| Node 1 | $\dfrac{3LP}{4H}$ | $\dfrac{P}{2}$ |
| Node 11 | $-\dfrac{3LP}{4H}$ | $\dfrac{P}{2}$ |

Table 19. Example 5: Element axial forces.

| Element # | Axial Force | Element # | Axial Force |
|---|---|---|---|
| 1 | $-\dfrac{PL}{4H}$ | 10 | $-\dfrac{LP}{4H}$ |
| 2 | $-\dfrac{P\sqrt{H^2+L^2}}{2H}$ | 11 | $-\dfrac{P\sqrt{H^2+L^2}}{2H}$ |
| 3 | $\dfrac{P}{4}$ | 12 | $\dfrac{P}{4}$ |
| 4 | $-\dfrac{P\sqrt{H^2+L^2}}{4H}$ | 13 | $-\dfrac{P\sqrt{H^2+L^2}}{4H}$ |
| 5 | $-\dfrac{LP}{4H}$ | 14 | $-\dfrac{LP}{4H}$ |
| 6 | $-\dfrac{P\sqrt{H^2+L^2}}{4H}$ | 15 | $-\dfrac{P\sqrt{H^2+L^2}}{4H}$ |
| 7 | $\dfrac{P}{4}$ | 16 | $\dfrac{P}{4}$ |
| 8 | $-\dfrac{P\sqrt{H^2+L^2}}{2H}$ | 17 | $-\dfrac{LP}{4H}$ |
| 9 | $-\dfrac{LP}{4H}$ | 18 | $-\dfrac{P\sqrt{H^2+L^2}}{2H}$ |

## 7 Validation of Results

The validation of any result is a critical step in establishing its reliability, especially for symbolic solutions, which are designed to be general and applicable to various configurations. Unlike numerical methods that directly produce specific results for a given set of inputs, symbolic solutions retain algebraic relationships, making them more versatile but also requiring rigorous validation to ensure accuracy across diverse scenarios. Validation is essential to confirm that the derived symbolic expressions accurately represent the behavior of the structural system for any configuration.

A systematic validation procedure was employed in this study. Using MATLAB's subs command, all symbolic variables were replaced with numeric values, allowing the symbolic results to be expressed numerically for direct comparison with other finite element software. This enabled precise evaluation of the symbolic results for specific truss configurations. Two software packages were utilized for this comparison: SAP2000 Ultimate (v21.2) and EngiLab Truss.2D 2022 Pro (v1.3). Additionally, for one example, comparisons were made with closed-form solutions from the literature.

### 7.1 Validation with SAP2000

All five truss examples were validated against SAP2000, with results matching perfectly for node displacements, support reactions, and element axial forces. Here, we present the validation for the third example, illustrated in Figure 5. The following properties were used for numerical calculations:

- Length (*L*): 8 m

- Height (*H*): 6 m

- Axial stiffness (*EA*): Derived using $E=200\times10^6$ kN/m² (200 GPa) and $A=4\times10^{-4}$ m² (4 cm²), resulting in $EA=8\times10^4$ kN.

- Point load (*P*): 100 kN

The truss was modeled in SAP2000, as depicted in Figure 8.

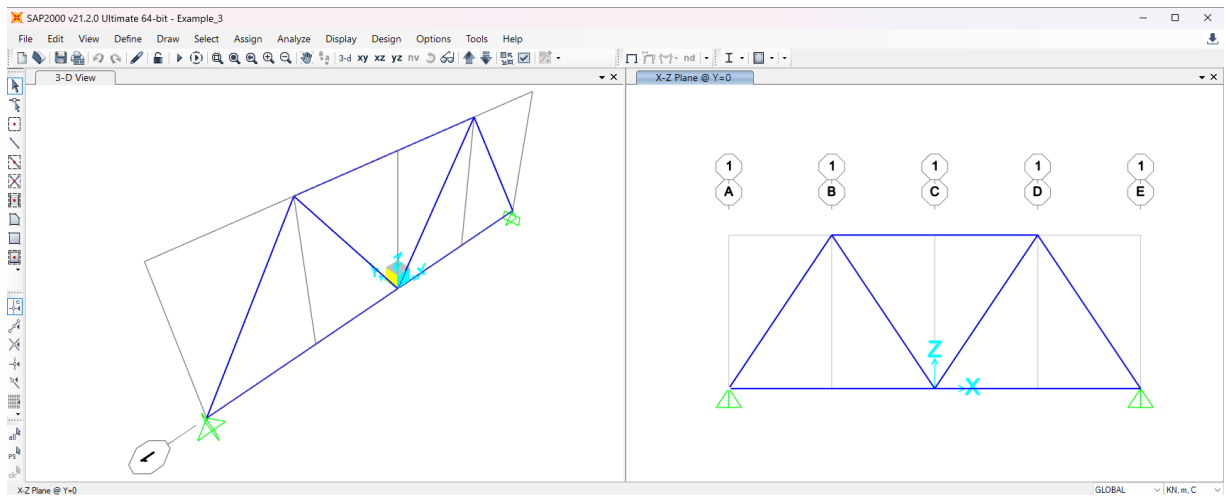

Figure 8. The truss model of Example 3, modeled with SAP2000.

Table 20 presents a comparison of element axial forces between the symbolic software and SAP2000. The results are identical, apart from minor differences in decimals due to rounding. This consistency extends to node displacements and support reactions, though those results are not shown here.

It is important to note that SAP2000 includes shear deformations by default in all frame and truss models. However, the Euler–Bernoulli beam theory, employed in our program, neglects the effects of shear deformation, as it assumes that transverse shear strain is insignificant. To ensure a meaningful comparison between the results from our symbolic program and SAP2000, the shear effects in SAP2000 were effectively minimized by setting the frame property modifier for shear area to a very high value ($10^6$). This adjustment aligns the SAP2000 model with the assumptions of the Euler–Bernoulli theory, allowing for consistent and accurate validation.

Table 20. Element axial forces for Example 3 – Comparison with SAP2000.

| Element # | Axial Force | Numeric value of symbolic expression | SAP2000 Result |
|---|---|---|---|
| 1 | $\dfrac{P}{2}$ | 50 | 50 |
| 2 | $-\dfrac{P}{2}$ | -50 | -50 |
| 3 | $-\dfrac{P(2H+L)}{2H}$ | -166.667 | -166.666 |
| 4 | $\dfrac{P\sqrt{4H^2+L^2}\cdot(H-L)}{2HL}$ | -30.0463 | -30.046 |
| 5 | $-\dfrac{P\sqrt{4H^2+L^2}}{2L}$ | -90.1388 | 90.138 |
| 6 | $\dfrac{P\sqrt{4H^2+L^2}}{2L}$ | 90.13878 | -90.139 |
| 7 | $-\dfrac{P\sqrt{4H^2+L^2}\cdot(H+L)}{2HL}$ | -210.324 | -210.323 |

## 7.2 Validation with EngiLab Truss.2D Pro

All five truss examples were also validated using EngiLab Truss.2D 2022 Pro, a specialized software for plane truss analysis. In every case, the results for node displacements, support reactions, and element axial forces matched perfectly. The validation for the fourth example, shown in Figure 6, is presented here. The following properties were used for numerical calculations:

- Length ($L$): 5 m

- Height ($H$): 6 m

- Axial stiffness ($EA$): Derived using $E$=200×10$^6$ kN/m² (200 GPa) and $A$=2×10$^{-3}$ m² (20 cm²), resulting in $EA$=4×10$^5$ kN.

- Point load ($P$): 50 kN

The truss was modeled in EngiLab Truss.2D Pro, as illustrated in Figure 9.

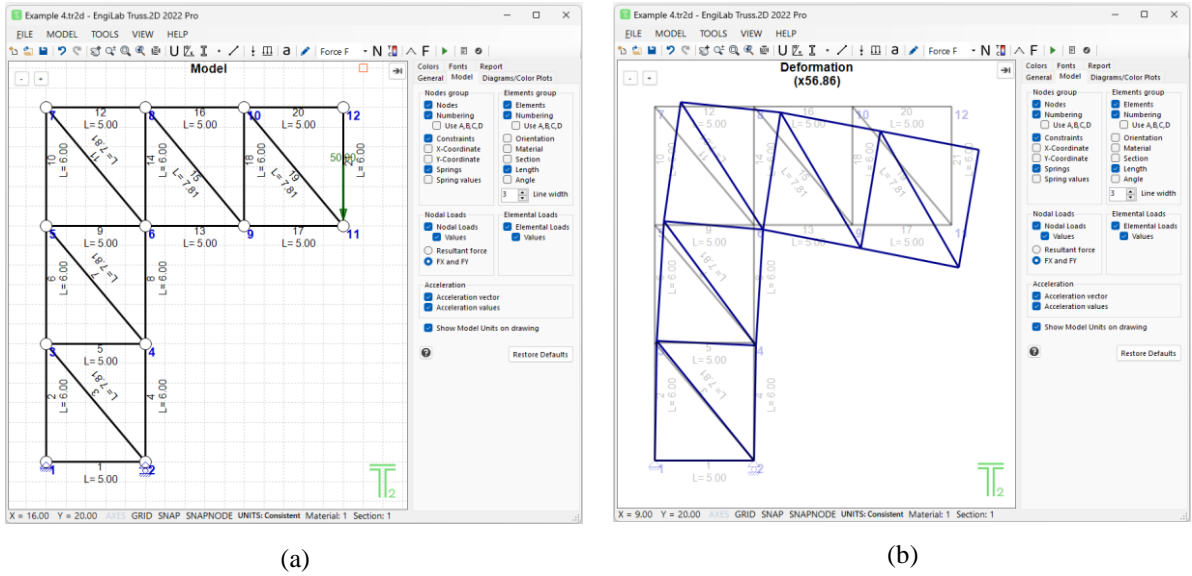

|     | (a) |     | (b) |
|-----|-----|-----|-----|

Figure 9. The truss of Example 4, modeled with EngiLab Truss.2D Pro:
(a) Truss model, (b) Deformed state of the model.

Table 21 compares node displacements between the symbolic software and EngiLab Truss.2D Pro, showing identical results. This consistency also applies to support reactions and element axial forces, though these results are not included here.

Table 21. Node displacements for Example 3 (for selected nodes) – Comparison with EngiLab Truss.2D Pro.

| Node # | Numeric value of symbolic expression | | EngiLab Truss.2D Pro Result | |
|--------|-------------------------|-------------------------|-------------------------|-------------------------|
|        | $x$-Displacement ($D_x$) | $y$-Displacement ($D_y$) | $x$-Displacement ($D_x$) | $y$-Displacement ($D_y$) |
| Node 6  | 0.0081   | -0.0045  | 0.0081  | -0.0045  |
| Node 7  | 0.02287  | 0.0045   | 0.02287 | 0.0045   |
| Node 9  | 0.007058 | -0.02095 | 0.00706 | -0.02095 |
| Node 11 | 0.006538 | -0.03827 | 0.00654 | -0.03827 |

## 7.3 Validation with results from the literature

The fifth truss example is based on a case from the literature, specifically Case No. 1 from Tinkov's work [38]. In this study, the author presents an analytical formula for calculating the vertical displacement of the central node (node 6 in our model). The formula is as follows:

$$\Delta_6 = \frac{P}{EF} \cdot \frac{Aa^3 + Bb^3 + Cc^3}{Db^2} \qquad (6)$$

Tinkov's methodology provides analytical solutions for displacements in various truss configurations, using the parameter $n$, which defines the truss topology and the number of repetitions of the main truss pattern, for each truss configuration. For the specific truss model shown in Figure 7, it is $n = 2$. According to the study [38], the values of the parameters $A$, $B$, $C$, and $D$ for the specific truss are as follows:

- $A = n+1 = 3$
- $B = n = 2$
- $C = \dfrac{n(n+1)(2n+1)}{3} = 10$
- $D = 2n^2 = 8$
- $a = L$ (the $L$ symbolic variable used in our study)
- $b = H$ (the $H$ symbolic variable used in our study)
- $c = \sqrt{a^2 + b^2} = \sqrt{L^2 + H^2}$

It is important to note that in Eq. (6) the variable $F$ represents the cross-sectional area. Substituting $F$ with our variable $A$ for the cross-sectional area, and taking the above values of the parameters into account, the final formula for the vertical displacement of node 6, as derived by Tinkov [38], becomes:

$$\Delta_6 = \frac{P}{EA} \cdot \frac{3L^3 + 2H^3 + 10\left(L^2 + H^2\right)^{3/2}}{8H^2} \qquad (7)$$

Eq. (7) matches the formula derived by our program for the vertical displacement of node 6, as presented in Table 17. The negative sign in the equation from Table 17 indicates that the displacement is directed downwards.

## 8   Sensitivity Analysis

Sensitivity analysis is an essential tool in structural engineering for evaluating how variations in input parameters influence a system's response. It is critical for assessing the robustness of a design and optimizing performance by identifying the parameters that most significantly impact structural behavior. Engineers often seek to understand how changes in material properties, geometric dimensions, or applied loads affect displacements, stresses, reaction forces, or internal forces. By offering insights into these relationships, sensitivity analysis facilitates more informed decision-making during the design process.

Symbolic solutions provide a distinct advantage for conducting sensitivity analysis. With symbolic representations of structural behavior, engineers gain access to closed-form expressions that explicitly describe how parameters—such as Young's modulus, cross-sectional area, or applied loads—affect the system's response. These expressions enable the direct calculation of partial derivatives of output quantities with respect to input parameters. For example, given a symbolic expression for displacement, one can compute the sensitivity of displacement at a specific point to changes in material stiffness or geometric properties. Similarly, partial derivatives of reactions or internal forces with respect to other parameters can be easily obtained.

For instance, in the third numerical example of Section 6.3, the vertical displacement at Node 2 is calculated symbolically (see Table 11) as follows

$$D_y = -\frac{P\left(2HL^2 + \dfrac{(4H^2 + L^2)^{3/2}}{2} + L^3\right)}{4EAH^2} \tag{8}$$

Using MATLAB's Symbolic Math Toolbox, partial derivatives of the displacement with respect to parameters such as axial stiffness ($EA$), length ($L$), or height ($H$) can be computed efficiently. Commands such as *diff* for symbolic differentiation and *simplify* for simplifying the resulting expressions make this process straightforward, yielding clear mathematical relationships that highlight the sensitivity of the system's response to design changes. The partial derivates of the displacement $D_y$ with respect to ($EA$), $L$ and $H$ can be easily found as:

$$\frac{\partial D_y}{\partial (EA)} = -\frac{P\left(2HL^2 + \dfrac{(4H^2 + L^2)^{3/2}}{2} + L^3\right)}{4(EA)^2 H^2} \tag{9}$$

$$\frac{\partial D_y}{\partial L} = -\frac{LP\left(8H + 6L + 3\sqrt{4H^2 + L^2}\right)}{8EAH^2} \tag{10}$$

$$\frac{\partial D_y}{\partial H} = \frac{P\left(2HL^2 + (4H^2 + L^2)^{3/2} - 6H^2\sqrt{4H^2 + L^2} + 2L^3\right)}{4EAH^3} \tag{11}$$

This symbolic approach offers a significant edge over numerical methods. In numerical analysis, results are typically confined to specific input values, requiring multiple reruns to evaluate parameter changes. Symbolic expressions, however, provide general solutions that inherently preserve the relationships between inputs and outputs, allowing for effortless exploration of parameter sensitivities without additional computational effort.

By enabling direct and efficient sensitivity evaluations, symbolic solutions empower engineers to gain deeper insights into how design modifications affect structural performance. This capability is especially valuable during design optimization, where small adjustments can greatly impact the efficiency and cost-effectiveness of a structure. Unlike numerical methods, symbolic analysis eliminates the need for repetitive computations, saving time and providing a clearer understanding of the system's behavior. This unique capability makes symbolic sensitivity analysis an indispensable tool for engineers.

## 9  Balancing Complexity and Practicality in Symbolic Computations

The symbolic implementation of MSA offers flexibility and insight but presents challenges in computational efficiency and scalability, especially for large or complex 2D truss systems. As the number of degrees of freedom, elements, or loading conditions increases, symbolic expressions for stiffness matrices, force vectors, and displacements can become overly large and computationally demanding. Addressing these challenges is critical to ensuring the practical applicability of symbolic MSA.

A key challenge in symbolic computation is the exponential growth of expressions as system complexity increases. While simple truss systems produce elegant and manageable symbolic solutions, larger models with many symbolic parameters often result in excessively lengthy and difficult-to-interpret expressions. Such complexity, even when technically valid, reduces the practicality of the symbolic approach. Compact and clear expressions are crucial for generating meaningful insights into structural behavior.

To address this, a hybrid symbolic-numerical approach proves effective and is fully supported by our program. This approach allows certain key parameters to remain symbolic while assigning numerical values to less critical variables. By combining symbolic flexibility with numerical efficiency, the analysis remains scalable and computationally manageable, enabling symbolic MSA to handle more complex 2D truss systems without excessive computational overhead.

Ultimately, the value of symbolic solutions lies in their clarity and utility. Compact, interpretable solutions enhance understanding and practical application, while excessively complex expressions undermine these benefits. Striking the right balance between symbolic and numerical methods ensures that the advantages of symbolic MSA are fully realized in 2D truss analysis.

## 10  Conclusions

This study introduces the development and application of an open-source MATLAB program for symbolic Matrix Structural Analysis of 2D trusses subjected to point loads. The freely available source code enables the efficient and precise generation of analytical solutions for plane trusses of any complexity. Beyond its practical engineering applications, the program serves as a valuable educational tool, providing clear and insightful symbolic solutions that enhance understanding of truss behavior.

The program extends its functionality beyond deriving analytical solutions for node displacements, support reactions, and element axial forces. It also facilitates sensitivity analysis by allowing users to compute partial derivatives of output parameters with respect to input variables, leveraging MATLAB's built-in symbolic differentiation commands. This capability is crucial for assessing how changes in design properties, such as material stiffness or geometric dimensions, influence structural response, making the tool invaluable for design exploration and performance evaluation.

While symbolic solutions offer significant advantages, it is essential to balance complexity and clarity. Although the program can generate detailed and complex symbolic expressions, overly intricate solutions may impede practical application and interpretation. Concise and actionable analytical expressions are preferable as they enhance usability and provide meaningful insights. Efficiency, scalability, and clarity remain key considerations when employing symbolic computations in structural analysis.

The program's capabilities have been demonstrated through various examples, showcasing its potential to deliver precise and insightful analytical solutions for 2D trusses. The results have been rigorously validated using three complementary approaches: comparisons with results from SAP2000 Ultimate (v21.2), EngiLab Truss.2D 2022 Pro (v1.3), and analytical solutions from the literature. These validations demonstrate perfect agreement between the program's outputs and the benchmark results, confirming the accuracy and reliability of the symbolic solutions and affirming the generality of the methodology.

A natural extension of this work would be to apply the methodology to other structural systems, such as 3D trusses, where additional complexities like three-dimensional geometry and nodal connectivity can be handled symbolically. This future direction holds great promise for further expanding the applicability of symbolic MSA and its contributions to structural engineering research and education.

## Conflict of Interest

The authors declare that the research was conducted in the absence of any commercial or financial relationships that could be construed as a potential conflict of interest.